\documentclass[11pt]{article}
\usepackage{fullpage}

\usepackage[utf8]{inputenc} 
\usepackage[T1]{fontenc}    
\usepackage{hyperref}       
\usepackage{url}            
\usepackage{booktabs}       
\usepackage{amsfonts}       
\usepackage{nicefrac}       
\usepackage{microtype}      
\usepackage{graphicx}
\usepackage{doi}

\usepackage{csquotes}
\usepackage{times}
\usepackage{latexsym}
\usepackage{setspace}
\usepackage{enumitem}
\usepackage{makecell}
\usepackage{xcolor}
\usepackage[round]{natbib}
\usepackage{authblk}

\newcommand\noi{\vskip .1in\noindent}

\newcommand{\ea}{EurekAlert! }
\newcommand{\eaa}{EurekAlert!}
\newcommand\parencite{\citep}
\newcommand\autocite{\citep}
\newcommand\textcite{\citet}
\renewcommand\cite{\citep}

\title{Linking Health News to Research Literature}

\author[1]{Jun Wang}
\author[2]{Bei Yu}
\affil[1]{Independent Researcher, Jamesville, NY 13078, USA}
\affil[2]{School of Information Studies, Syracuse University}

\date{}

\begin{document}
\maketitle

\abstract{
Accurately linking news articles to scientific research works is a critical component in a number of applications, such as measuring the social impact of a research work and detecting inaccuracies or distortions in science news.  Although the lack of links between news and literature has been a challenge in these applications, it is a relatively unexplored research problem.  In this paper we designed and evaluated a new approach that consists of (1) augmenting latest named-entity recognition techniques to extract various metadata, and (2) designing a new elastic search engine that can facilitate the use of enriched metadata queries.  To evaluate our approach, we constructed two datasets of paired news articles and research papers: one is used for training models to extract metadata, and the other for evaluation.  Our experiments showed that the new approach performed significantly better than a baseline approach used by altmetric.com (0.89 vs 0.32 in terms of top-1 accuracy).  To further demonstrate the effectiveness of the approach, we also conducted a study on 37,600 health-related press releases published on EurekAlert!, which showed that our approach was able to identify the corresponding research papers with a top-1 accuracy of at least 0.97. 
}


\section{Introduction}

Accurately linking news articles to scientific research works is a critical
component in a number of research areas. One of them is altmetrics
\parencite{sud2014evaluating}, which is to measure the social impact of a research work
based on its mentions in news and social media. Another example is
evaluating information quality in science communication, e.g., detecting if a
news article misinterprets a research work, such as making causal
claims from correlational findings, or inference to humans from animal
study results \parencite{sumner2014association,yu2020exaggerationCOLING}.

Although research in these areas requires linking news articles to research papers,
the links are missing in a large number of science and health news
reports, since including links is not a standard practice for traditional news
outlets \parencite{liu2013five}. Moreover, the existing links are often outdated or pointed to the wrong sources.
For example, based on our calculation, more than 25\% of the links included in
Reuters news articles are outdated, with many of them pointing to the journals'
homepage, which often shows the latest publications instead of the specific
research paper cited in the news article.

The problem of missing and wrong
links calls for an automated solution to match news articles with
the corresponding research papers \parencite{liu2013five,ravenscroft2018harrigt}.
However, linking news to literature is still a relatively unexplored research
area. To our best knowledge, \parencite{ravenscroft2018harrigt} is the only published
study that tackled the exact problem. They developed a system named HarriGT to
link news articles to research papers. In addition, altmetric.com, arguably the
most widely used altmetrics tool, briefly described their proprietary
techniques for picking up mentions of research works in news \parencite{liu2013five}. 
Both studies modeled the linking problem as an ad hoc information retrieval task 
with a two-stage approach: first,
use named-entity recognition (NER) or text mining techniques to extract some metadata items, such as
author and affiliation names, from a news article; second, use the metadata to
query research literature databases to find the corresponding research paper. 
The performance of altmetric.com is not disclosed. 
HarriGT reported a top-1 accuracy of 0.59, indicating the need for improvement toward a deployable tool. 

The above approaches have only utilized a limited number of extracted metadata and
off-the-shelf search engines. Given that the advancement of natural language processing
and information retrieval offers new ideas toward solving the linking problem, 
in this study we explore the following two questions: (1) Can utilizing more
metadata and content information improve the accuracy of linking? (2) Can we design a
new literature search engine to achieve better linking performance than current literature
search tools like the CrossRef used by altmetric.com?  

This study seeks answers to the above research questions by designing and evaluating a
new two-stage approach: (1) develop a new NER model to extract more metadata such as
journal names from news articles, (2) develop a new search engine to
incorporate rich metadata and content information into the search strategy. 

Focusing on the health domain, we evaluated this approach using a test data set 
comprised of news articles from various outlets. We choose the health domain for two reasons. 
First, health is one of the most popular topics in
science news, which often gets more coverage than other science topics. 
For example, 40\% of the articles published on 
EurekAlert!---the major platform for universities and journal publishers to 
distribute press releases to journalists and the public---are under the category of {\em Medicine \& Health}.
Research has also suggested that studies in the domain of health and medicine
account for the majority of media coverage \autocite{suleski2010scientists}.
Second, the PubMed corpus
serves as an ideal evaluation data set for this task:
it has the largest freely-available database of publications in the health and medicine domain
that contains rich metadata items annotated by the National Library of Medicine.

To demonstrate the effectiveness of our approach, we further present a case study in
which our linking system was able to recover the links between 40,000
medicine and health research-related press releases on \ea and their associated
research articles in PubMed with 98.1\% precision and 86.9\% recall. Since press releases have
now become the dominant link between academia and mainstream news media
\cite{sumner2014association, brechman2009lost}, linking research papers with
press releases will be a valuable tool for studying science communication.

%
\section{Related Works}

HarriGT is the only published system on linking news to science literature
\autocite{ravenscroft2018harrigt}. The authors developed a corpus of about 300 news
articles and applied general-purpose NER techniques to recognize personal
names and institutions in the news articles. Then named entities-based
queries were sent to literature search engines (Microsoft
Academic, Scopus, and Springer) to find research papers published within $\pm$90 days.
The search results were further re-ranked to ensure the most relevant papers 
are at the top of the search result.
The authors reported a top-1 accuracy of 0.19 (before re-ranking) or 0.59 (after re-ranking),
 indicating a need for improvement toward a deployable tool.

Altmetric team ({\em altmetric.com}---arguably the most widely used altmetrics tool)
developed a text mining system to extract journal titles and author names
from news articles and then used the information to query CrossRef (\url{https://www.crossref.org/})
with a time window of $\pm$45 days from the date of news release \autocite{altmetricAboutourdata,maclaughlin2018predicting}.
Neither algorithm details nor performance has been disclosed to the public,
although their brief report admitted room for improvement \autocite{liu2013five}.

The above approaches have a number of limitations. First, although general-purpose
NER techniques \autocite{lample2016neural} can recognize some metadata items such as
authors and affiliations as named entities ({\em persons} and {\em organizations}),
specialized NER models are needed for identifying more metadata items that are not targets
of common NER tools, such as journal names. 
To some extent, the task of extracting a journal name and using it as a link between news articles and research papers
is similar to another specialized NER task---identifying gene and protein names 
from biomedical literature and then linking research papers with gene and protein names 
\autocite{hoffmann2005ihop, garten2010recent}.

Second, the current search strategy
only includes a limited number of metadata items in queries, such as author and affiliation in
\autocite{ravenscroft2018harrigt}, while the usefulness of rich metadata and content
information has not been investigated. 
To some extent, locating research papers associated with a news article can be considered as a navigational search task.
Prior user studies on navigational search have shown that compared to exploratory search, 
navigational search \autocite{jansen2008determining} might benefit 
from longer queries \autocite{athukorala2016exploratory}. 
Hence, it is worth exploring whether expanding queries with more metadata would help link news articles to research papers.

Third, existing literature search engines such as CrossRef, 
have limitations when used for linking news articles to research papers. 
Here we list three of them.
(1) 
Given a query with two or more metadata items,
none of the current literature search engines supports the relevance scoring function
that can not only sum up the individual scores calculated from each query component,
but also weight them to reflect the importance of certain metadata.
(2)
Given a news article, one would expect that the publication date of its corresponding paper 
should be as close as possible to the news release date.
However, current search engines only allow one to specify a date range (e.g., $news\_date\pm90 days$)
which treats papers inside the range equally and makes papers outside the range impossible to retrieve.
(3)
Off-the-shelf search engine ranking algorithms give higher author match score to candidate research papers that have fewer authors
than those that have more authors.
However, health and medicine research often involves large teams of authors. 
According to the Nature Index, the average number of authors listed on a life science paper is 13.7 in 2016 \cite{natureindex2018}.

These limitations call for more research on better approaches for linking news articles to research papers.

%
\section{Problem Modeling}
Similar to prior studies, we also decomposed the linking problem into two
tasks: metadata extraction and information retrieval. We proposed and tested the
following new ideas.

For the metadata extraction task, we trained a specific NER model for journal name
identification. Before sending sentences to this NER model, a filtering strategy was first used to 
predict whether a sentence contains a journal name. 
Thus, only the ``journal sentences" that mentioned journal names will be sent to 
the journal name identification model. Existing NER tools
were used to identify author names and affiliations, as previously done in
\autocite{ravenscroft2018harrigt}.

For the information retrieval task, we developed a new literature search engine that uses
elastic search, which not only allows searching by weighting multiple metadata items,
but also incorporates date proximity into a decay function so that
research papers published closer to the news release date are ranked as more
relevant. Given that news articles contain 
more content---such as research background, methods, and results---than just the above mentioned metadata,
we also integrated the news content into our search strategy.

\section{Datasets}

Two news data sets were created to train and evaluate the above mentioned 
models for sentence filtering and metadata extraction. 
First, a corpus created from the {\em Reuters} health news archive
(\url{https://www.reuters.com/news/archive/healthNews})
was used as the dataset to train the models.
Second, to ensure the models' generalizability to other news sources,
another data set, {\em Cardiff}, containing news articles from a broad range
of sources, was created to evaluate the trained models.
The Cardiff dataset was also used to evaluate our PubMed corpus-based elastic search engine.
Our data is available at \url{https://github.com/junwang4/linking-health-news-to-pubmed}.

\subsection{Training dataset: Reuters health news}
We chose Reuters as the news source for developing our training dataset because it has 
a large accessible archive of health news, including 8,600 news articles 
which contained one or more bitly urls linking to related research publications. 
Using the bitly urls, we can identify the DOI links of the journal papers through fetching and parsing the linked web pages.
During the procedure of DOI identification,
however, we found that not all of the bitly urls are reliable:
more than a quarter of the links were outdated or redirected 
to the entrance page of a journal website, 
which often features the latest journal issue 
instead of the specific research paper cited by the news article. 
Additionally, for simplicity, in this paper we focused on news articles 
reporting on a single research work, namely, those with only one bitly url.
These news articles usually report on the latest findings from individual studies;
however, some news articles report on multiple studies, 
especially when introducing a research area rather than an individual study. 
It will be our future work to study the case of one news article citing multiple research works.
In the end, we obtained 5,533 valid pairs of health news articles and research papers. 


\subsection{Test dataset: Cardiff}

In 2014, \textcite{sumner2014association} at Cardiff University and two other institutes, 
published a study on how the exaggeration in health medical news of UK could be influenced by
academic press releases. They collected hundreds of health-related press releases 
alongside associated research papers and news stories, and made them publicly available at
\url{https://github.com/liampshaw/pr\_hype}.
From this source we obtained a dataset of 604 news articles and 215 corresponding research papers.
This dataset contains a total of 21 news sources, 
with the top 3 (BBC News, Daily Mail, and Telegraph) accounting for 40\% of the news.

\begin{figure}
\center
\begin{tabular}{c@{}c}
\includegraphics[width=0.49\columnwidth]{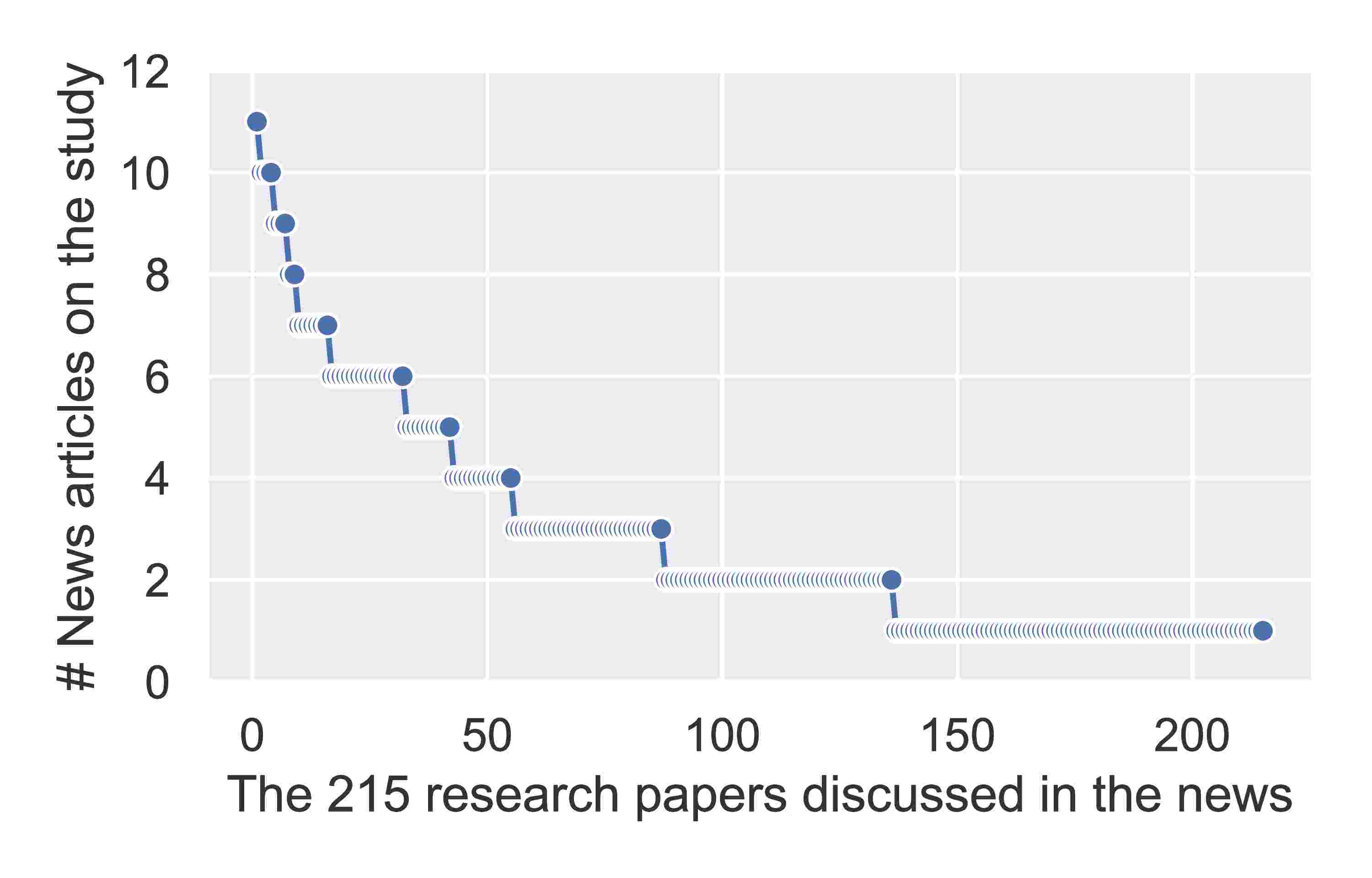}
&
\includegraphics[width=0.49\columnwidth]{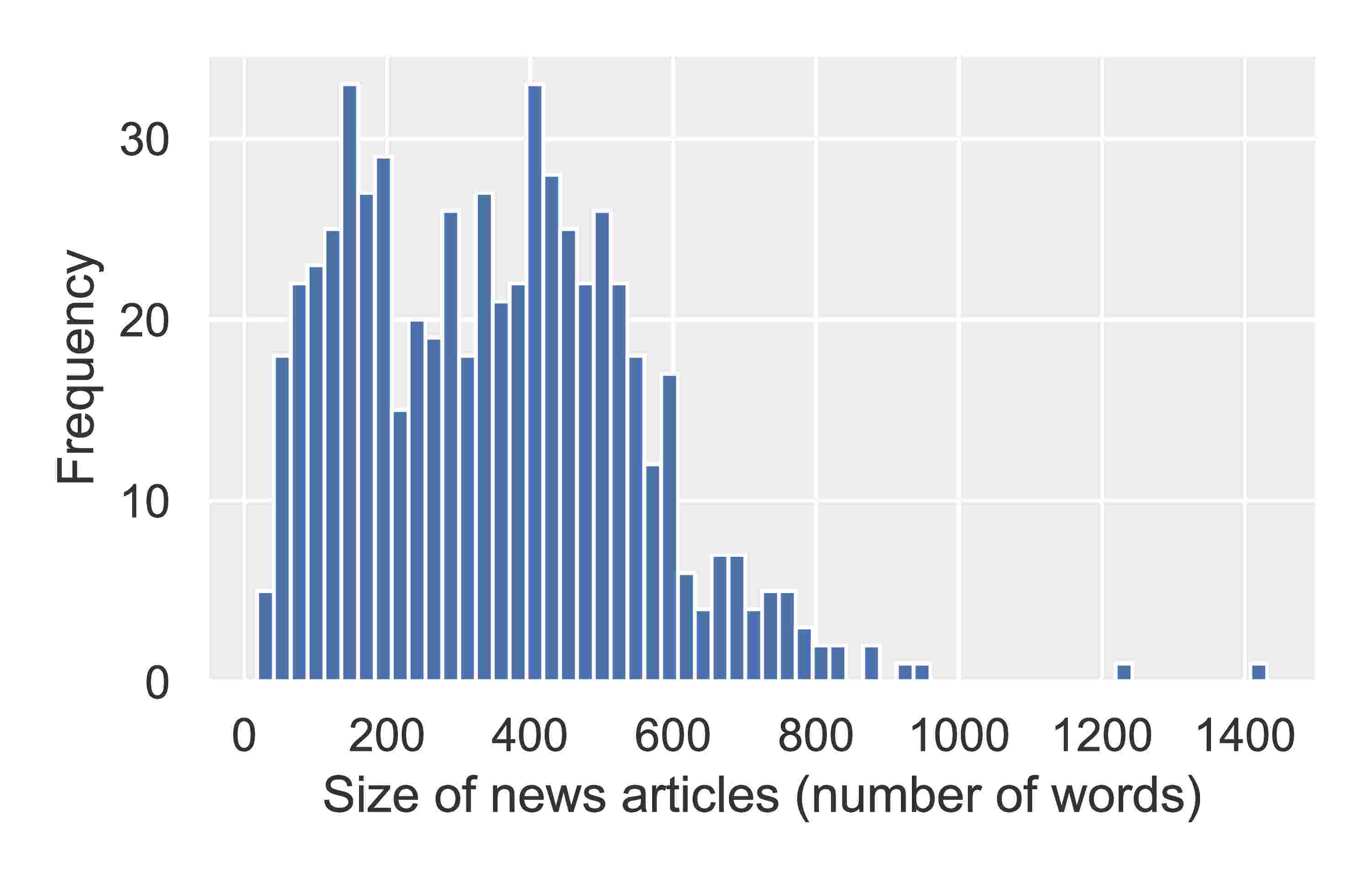}
\\
({\bf a}) Number of news agencies picking up a study
& 
({\bf b})  Distribution of the size of news articles
\end{tabular}
\caption{Properties of the Cardiff dataset.}
\label{figSumnerCnt}
\end{figure}

The Cardiff dataset presents a more diversified news sources than the Reuters dataset.
First,
as shown in Fig. 1(a), over 60\% of the research works (135 out of 215) were picked up
by two or more news agencies. Second, as shown in Fig. 1(b), there are two modes
in the distribution of the length of the news articles: one is about 150 words and
the other 400 words. 
In contrast, the Reuters news articles follow the normal distribution with a mean around 600 words. 
Since the training data and the test data came from different news sources and writing styles, 
a good test performance can ensure the model trained on one news outlet is generalizable to others.

\section{Metadata Extraction}
\label{sectJournal}

\subsection{Journal name identification}

\begin{figure}
\center
\includegraphics[width=.95\textwidth]{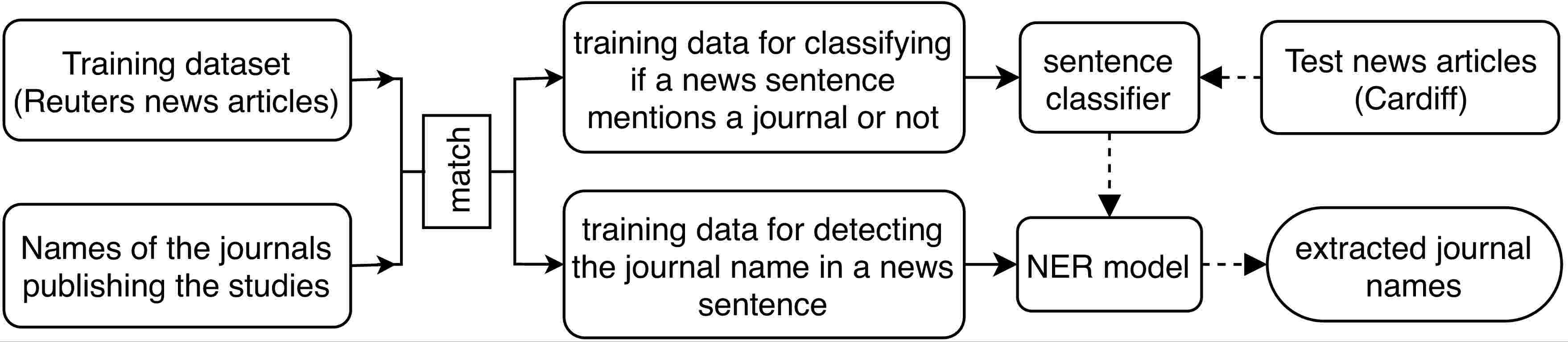}
\caption{Framework of annotating, training, and recognizing journal names in news articles.}
\label{figTrainFramework}
\end{figure}

Since journal name identification is not included in standard NER models, 
we developed a new method to extract journal names from news articles.
The new method includes two steps: 
first, develop a {\em journal sentence filter} 
to remove sentences that did not mention any journal; 
second, develop a specialized NER model to extract journal names 
from the sentences that did mention a journal.  
See Fig.~\ref{figTrainFramework} for an illustration of the whole procedure.

To develop the {\em journal sentence filter},
we first built a training data set 
consisting of sentences that mention a journal (i.e. {\em journal sentences})
and sentences that do not (i.e. {\em non-journal sentences}). 
For each pair of linked news article and research paper in the Reuters corpus,
since we have the name of the journal that publishes the research paper,
we can locate the specific journal sentence that mentions the journal name in the news article.
Using this method, from the above mentioned 5,533 pairs of news articles and research papers,
we were able to locate 4,876 journal sentences.
For the remaining 657 (12\%) news articles, the original
journal names were either not mentioned or mentioned with an alternative name, 
such as ``PNAS"
or ``Proc Natl Acad Sci'' for ``Proceedings of the National Academy of Sciences of the United States of America''.
Regardless of mentioning original journal names or alternative names,
all journal sentences share common linguistic patterns when
referring to a journal (e.g. many journal sentences contain such
words as {\em journal, published, report,} or {\em write}); therefore, our journal sentence
prediction model should be generalizable to all journal sentences.

We then randomly sampled from all news articles another set of
4,876 sentences as non-journal sentences. 
This balanced dataset was used to train a journal sentence classification model.
We compared the performance of two text classification methods:
Linear SVM (count vectorizer with 1,2,3-grams, which is better than tf-idf vectorizer)
and BERT (case-based).
To have a fair evaluation,
we adopted a group-based training-test set split method: the journal sentences with the same journal name
should appear either in the training or in the test set.
For LinearSVM, the classification accuracy is 98.6\%,
and for BERT, it is 99.7\%. Hence we adopted the BERT method in our system.

We also used the BERT-based pretrained model to fine-tune a journal NER model. 
On the 4,876 journal sentences, via 5-fold cross-validation, 
our fine-tuned journal NER model can obtain an accuracy of 99.9\% and F1 of 0.989 (both precision and recall are 0.989)
at the level of full named entity \cite{sang2018conll}.
In comparison, a baseline method that uses the {\em Stanza} NER to extract ORG entities as journal names
from the journal sentences resulted in an accuracy of 94.2\% and F1 of 0.378 \cite{sang2018conll}.

\subsection{Author name and affiliation identification}

For author name extraction,
we adopt a simple approach by extracting personal names
through the use of the above mentioned latest NLP package {\em Stanza}.
The matured NER techniques ensure a high recall performance, 
although the precision may be affected by
the cases where a personal name is not an author.
For this task, recall is more important than precision in that
if an author is misclassified as a non-author and thus not included in the query, 
the corresponding research paper would be difficult to retrieve.
On the other hand, the occasional inclusion of non-author names in the queries
does not affect the performance of the BM25 algorithm used in the Elastic search engine (see next section).  
Therefore, we decided to take this simple approach of assuming all recognized personal names as the authors.

For similar reason, we took a similar approach for author affiliation extraction
by using the Stanza NER module to extract organization names as author affiliations.

\section{Experiments with Search Engines}
\label{sectSearchEngine}

The above extracted metadata constitute the basis for constructing search queries. 
However, it is unclear which metadata items are more important, 
and what combination of the metadata items 
---including journal, author names, affiliations, and even news title--- 
is the most useful for query construction.

In this section we first present 
the evaluation result of 
using various metadata combinations to create queries for CrossRef,
the literature search engine used by altmetric.com.
The result shows that such
off-the-shelf literature search engines lack the capacity to utilize rich metadata.
We then describe the design and evaluation of a new search engine
that can fully harness various metadata.

\subsection{Experiments with CrossRef}

CrossRef is a major DOI registration agency that has
registered more than 120 million works \cite{crossref2021cache}.
CrossRef provides a date range filter
in the form of {\em from\_create\_date} and {\em until\_create\_date}. 
Like altmetric.com who adopted CrossRef, we also used a
window of $\pm 45$ days from the date of news release, to set up a query.
For the remaining part of the query, it is constructed
with various combinations of author name, affiliations, journal
name, and/or news title. After the query is created, it is sent to CrossRef via its API.
For example, if we want to query CrossRef with all four metadata, the request will look like:
``https://api.crossref.org/works?query.author=[news\_author]\&query.container-title=[news\_journal]\&query.affiliation=[news\_aff]\&query.bibliographic=[news\_title]''.

Like \autocite{ravenscroft2018harrigt}, 
we also used {\em top-k accuracy} to measure performance.

\noi {\bf Definition of top-k accuracy.}
For a news article, if its associated paper is in the top-k list
of the retrieved results, we count it as a {\em top-k hit}. For $n$ news articles, if
there are $m$ top-k hits, we say the top-k accuracy is $\frac{m}{n}$.

Note that we chose to use top-k accuracy instead of other measurements such as 
nDCG (normalized discounted cumulative gain) 
because in our evaluation setting, as mentioned above,
our dataset only includes those news articles that discuss one research work.

\begin{table}[!h]
	\center
\caption{ Retrieval performance with CrossRef.}
\begin{tabular}{@{ }lccc@{ }} \toprule
Experiment & Top-1 & Top-3 & Top-5 \\ \midrule
Au & 0.121 & 0.194 & 0.227 \\
{\bf AuJo} &{\bf 0.320} &{\bf 0.416} &{\bf 0.450} \\
AuAf & 0.038 & 0.048 & 0.050 \\
AuJoAf & 0.293 & 0.373 & 0.397 \\
AuJoTi & 0.288 & 0.351 & 0.363 \\
AuJoAfTi & 0.262 & 0.316 & 0.323 \\
\midrule
\multicolumn{4}{@{}l}{Au: Author; Jo: Journal; Af: Affiliation; Ti: Title}
\\
\bottomrule
\end{tabular}
\label{tabCrossrefResult}
\end{table}

Table~\ref{tabCrossrefResult} shows the results of running experiments with various combinations from author name
only  (Au) to the complicated combination of author name, journal name,
affiliations, and news title (AuJoAfTi). As shown in the table, AuJo, the
combination of author name and journal name, performs the best. This
counter-intuitive result can be explained by the following search strategy used in CrossRef.
Suppose a query involves two fields (author and journal),
and is expressed in the following form:
``https://api.crossref.org/works?query.author=[news\_author]\&query.container-title=[news\_journal]''.
When the CrossRef engine receives the query, the engine only returns those documents whose author names have common tokens with [news\_author]
and whose journal name has common tokens with [news\_journal].
In other words, when there are two or more query fields,
CrossRef engine interprets their relationship as logical {\em AND}.
A desirable search engine should treat the relationship among multi query fields as a kind of {\em OR} or {\em SUM}.
The elastic search engine introduced below has this desirable property
that allows us to harness the true power of a variety of metadata.

\subsection{Experiments with Elastic search}

To overcome the limitations of CrossRef, we built a new literature search
engine based on Elastic search, which is the most popular enterprise search
engine that is based on the open-source Apache Lucene search engine library.

\noi{\bf Building search engine index.}
For this study we chose to use PubMed metadata (\url{https://ftp.ncbi.nlm.nih.gov/pubmed/}) to build our database index.
PubMed is the largest biomedical literature database with over 20 million records listed with abstracts.
It contains the most complete metadata items, including journal name, author names and
affiliations, paper title, abstract, and publication date (year, month, and day).
Note that each paper may have up to four types of dates: journal publication date, PubMed publication date, online publication date, and accepted date. 
We assume the release date of a news article is around the earliest publication
date, when the paper is available for public access.  
It is usually the online publication date. When it's not available, the earlier
one between the journal publication date and PubMed publication date is used as
the earliest publication date. 
Sometimes an error occurs such that the earliest date appears to be even
earlier than the accepted date. For example, when the publication month/day of
a paper are missing they are simply recorded in PubMed as the first day of the
year. In these cases, the accepted date will be taken as the earliest date.

\noi{\bf Developing a specialized elastic search engine.}
The default relevance scoring or ranking algorithm used in the elastic search is BM25.
Given a query Q, containing keywords $q_{1},\cdots,q_{n}$, the BM25 score of a document D is
\begin{equation} 
	\label{eq:bm25} 
	\textrm{score}(D,Q) = \sum_{i=1}^{n} \textrm{IDF}(q_i) \cdot
	\frac{f(q_i, D) \cdot (k_1 + 1)}{f(q_i, D) + k_1 \cdot (1 - b + b \cdot \frac{|D|}{\textrm{avgdl}})} 
\end{equation}
where $f(q_{i},D)$ is $q_{i}$'s term
frequency in document D, $|D|$ is the length (number of words) of document D,
\textrm{avgdl} is the average document length of a corpus,
and $\textrm{IDF}(q_i)$ is the inverse document frequency of the query term $q_{i}$.
In the above formula,
$k_1$ is a parameter (whose default value is 1.2) to control how ``terms occurring extra times add extra score'',
and $b$ is a parameter (whose default value is 0.75) to control how document length affects similarity score.

A metadata-rich query is a combination of multiple metadata subqueries.
For example,
an author subquery $Q\_au$ is all personal names extracted in a news article. 
Correspondingly, an author document $D\_au$ is the names of all authors of the document D.
Given a query with five subqueries \{au, jo, af, ti, co\} ({\em co} indicates news \underline{co}ntent, details later),
the overall relevance score between a news article and a research paper document is
\begin{equation} 
	\label{eqScoreOverall} 
	\textrm{weighted\_score}(au, jo, af, ti, co) = 
\sum_{k \in \{{au, jo, af, ti, co} \} } weight_k \cdot \textrm{score}(D_k, Q_k)
\end{equation}

\noi{\bf Four design features.}
Having a home-made search engine allows us to take advantage of  the elastic engine's  flexibility.
Here we presented four features of the new search engine. The performance gain of using these features is presented in 
Table~\ref{tabElasticResultFourFeatures}.

\begin{enumerate}[noitemsep,leftmargin=*]
	\item Because journals may have alternative names or abbreviations, when creating indexes for
the search engine, we augmented the journal names registered in PubMed with a
list of alternative names, obtained through linking the journal to the NLM medical journal
name database via the journal's ISSN \cite{NLMcatalog}.

\item Since elastic relevance scoring algorithm allows a weighted linear combination of individual scores to be calculated from subqueries,
we can conduct a grid search to find an optimal weight setting.
For example, we found that in an optimal weight setting,
the author subquery is less important than the journal subquery.

\item In the above BM25 score formula, parameter $b$ is used to control 
	how document length affects similarity score.
	By default, $b=0.75$, which makes the ranking algorithm give higher score to candidate research papers that have fewer tokens than those that have more tokens.
When used in calculating match score for the author subquery, this causes a problem that the algorithm would favor papers with fewer authors despite that health research papers often involve many authors. For example,
when a news article mentions two personal names: one is an author of the reported research paper that has multiple authors,
and the other is a journal editor who published an editorial in the same journal commenting on the reported work. 
The ranking algorithm with the default $b$ value would pick the editorial as a better match than the research paper since the editorial has only one author. 
In our experiments, we found that the smaller the $b$, the better the performance. 
Hence we set $b=0$ for author subquery.
As shown in Table~\ref{tabElasticResultFourFeatures}, using this feature alone resulted in a 40\% performance gain in terms of top-1 accuracy (from 0.374 to 0.523).

\item Elastic search supports a type of decay function that can decay relevance
score depending on how far a value is from a given origin. 
It can be particularly helpful for finding research papers whose publication
dates are expected to be as close as possible to the date of news release.
In our design, the date decay function takes the following form.
$$\textrm{date\_score}(paper\_date, news\_date) = e^{\log(0.5)/180 \cdot max(0, |paper\_date-news\_date|-7)}, $$
in which 7 indicates that if the date difference is within 7 days, the match score will be 1 (perfect match);
and (0.5, 180) indicates that if the date difference is 180 days (speaking precisely, 180+7 days), the score will be 0.5.

To integrate the date decay score (which ranges from 0 to 1) into the above weighted score (Eq.~\ref{eqScoreOverall}),
we used the following default score combination approach provided by Elasticsearch:
$$
\textrm{final\_score\_overall} = 
\textrm{date\_score}(paper\_date, news\_date) 
\times
\textrm{weighted\_score}(au, jo, af, ti, co)
$$

When taking the above decay function-based date similarity score into the overall score formula,
those research papers whose publication date is close to the news release date will be ranked higher.
In contrast, if using a traditional date matching method (like existing literature search engines),
the date match score is binary: either within a window or not.
\end{enumerate}

\begin{table}[h]
	\center
	\caption{Effects of using various features for elastic search.
	(For comparison with CrossRef, only author and journal fields are considered.)}
\begin{tabular}{@{ }llcccc@{ }} \toprule
	Experiment &	& Top-1 & Top-3 & Top-5 & Average time (seconds) \\ \midrule
	Baseline (AuJo) & & 0.374 & 0.487 & 0.530 & 0.03 \\
	+ Alternative journal names & & 0.429 & 0.555 & 0.576 & 0.03 \\
	+ Setting an optimal weight for subqueries &	 & 0.434 & 0.548 & 0.570 & 0.03 \\
	+ Setting BM25 parameter $b=0$		&	 & 0.523 & 0.647 & 0.669 & 0.03 \\
	+ Decay function-based date scoring &	 & 0.409 & 0.531 & 0.571 & 0.08 \\
	With all four features:								&	 & 0.644 & 0.740 & 0.757 & 0.08 \\
\bottomrule
\end{tabular}
\label{tabElasticResultFourFeatures}
\end{table}

\noi {\bf Effects of using enriched metadata.}
Table~\ref{tabElasticResult} shows how the use of additional metadata (news title, author affiliations, and news content)
in the search query affects the retrieval performance. 
For the news content subquery, we only considered the first 300 tokens in a news body --- our experiments show that 
the first 300 tokens are good enough to achieve an optimal performance. 
Note that the results shown in the table were obtained by using all four features described above.
Specifically, for the feature of weight setting, we ran a grid search for optimal weights of the five types of subqueries 
(Author, Journal, Affiliation, Title, and Content),
and obtained the following setting:
Au/1, Jo/1.5, Af/0.3, Ti/0.3, and Co/0.2.

As shown in the table, 
adding news title, author affiliation, or news content into search queries can improve retrieval performance,
and the largest performance boost comes from using the news content in a query.
This boost is made possible by adding to the search engine index a text field that consists of all relevant 
information, including paper title, author names and affiliations, journal name, and paper abstract. 

Table~\ref{tabElasticResult} also displays the time needed to run each experiment
in terms of a single search task. The experiment of using news
content as part of a query takes almost 5 times as much time as
the one without the use of news content, suggesting there
is a considerable tradeoff between linking accuracy and response time.

\begin{table}[h]
	\center
\caption{Effects of using additional metadata in the search query.
}
\begin{tabular}{@{ }llcccc@{ }} \toprule
	&	Experiment & Top-1 & Top-3 & Top-5 & Average time (seconds) \\ \midrule
	1 & 	AuJo & 0.644 & 0.740 & 0.757 & 0.08 \\
	2 & 	AuJoTi& 0.689 & 0.785 & 0.791 & 0.10 \\
	3 & 	AuJoAf& 0.699 & 0.788 & 0.811 & 0.13 \\
	4 & 	AuJoCo& 0.879 & 0.927 & 0.940 & 0.38 \\
	5 & 	AuJoAfTiCo& 0.892 & 0.934 & 0.947 & 0.42 \\ 
\bottomrule
\end{tabular}
\label{tabElasticResult}
\end{table}

\section{An Application: Linking Press Releases with Corresponding Research Articles}

Nowadays, newspapers rely heavily on press release materials to write science news stories 
\cite{semir1998press, schwitzer2008us, taylor2015medical}. 
However, an institutional press release serves the dual purpose of responsible science reporting 
and marketing \cite{carver2014public, caulfield2015commercialization, samuel2017uk}.
With press releases becoming the dominant link between academia and news media,
concerns have also increased regarding exaggeration in press releases, 
such as reporting correlational findings as causal and extrapolating results from animal studies 
to humans \cite{sumner2014association,li2017nlp,yu2020exaggerationCOLING}.

To identify and correct exaggeration in press releases, the first step is to link press releases 
to the original research papers, 
and then different versions of claims can be compared to determine whether overstatements 
occurred and how to correct them \cite{yu2019EMNLPCausalLanguage,yu2020exaggerationCOLING}.
However, linking press releases to the original papers is not a trivial task.
For example, \ea has published about 146,000 health-related press releases since 1996, 
and only about 40,000 (27\%) of them have DOI links.

We use the press releases with known DOI links as a test data set to further evaluate our linking approach.
Among the 40,000 press releases with known DOI links, 37,600 or 94\% are registered in PubMed. These 37,600 press releases and the corresponding PubMed papers are then used to test our linking approach. The main reasons that a health-related DOI is not found in PubMed include (1) a non-health journal published a health-related paper, e.g. some papers on {\em Physics of Fluids} are about COVID-19; (2) new health journals have not been indexed by PubMed due to the two-year publication requirement, e.g. {\em Nature Cancer} was established last year;
\footnote{Some papers published in ``Physics of Fluids'' or ``Nature Cancer'' can still be found in PubMed database due to the fact that the fulltext of those papers, as author manuscripts, were deposited in PubMed Central (PMC) in compliance with public access policies.} 
 (3) typos in DOI links.

The linking performance for the 37,600 press releases is reported in (Table~\ref{tabEAperformance}), 
which shows a highly accurate matching result, with the top-1 accuracy at 97.2\%.
\begin{table}[!h]
	\center
\caption{Performance on 37,600 DOI-known \ea press releases.}
\begin{tabular}{lccccc} \toprule
	Accuracy & Top-1 & Top-2 & Top-3 & Top-5 \\ \midrule
		& 0.972 & 0.984 & 0.987 & 0.989 \\ \bottomrule
\end{tabular}
\label{tabEAperformance}
\end{table}

\noi {\bf Mismatch analysis.}
To have a better understanding of the mismatched cases, we randomly sampled 25 such cases, and found that
\begin{enumerate}

\item For 10 of them, two papers were reported (for example, ``in two new studies, both published in Science Immunology...''),
	but only one of the two links was given, because only one link was allowed in EurekAlert!'s webpage section of {\em Related Journal Article}.
	Despite that, our matching system was able to rank all the $10\times 2$ papers within top 2.

\item For 2 of them, our results were actually correct; incorrect DOI references that link to completely unrelated papers were posted to \eaa.

\item For 13 of them, our top-1 matches were indeed incorrect.
The most challenge cases are the studies that were published online long time before the press release date, e.g. over a year ago.
	
\end{enumerate}

This error analysis shows that if taking into consideration that our approach works well on 
about half of the mismatched cases, the actual top-1 accuracy should be over 98\%.
With this level of high accuracy, we are confident in applying the approach to the remaining \ea press releases that are not yet linked.

\begin{figure}[!t]
\center
\includegraphics[width=.55\textwidth, trim={0 0 0 0cm},clip]{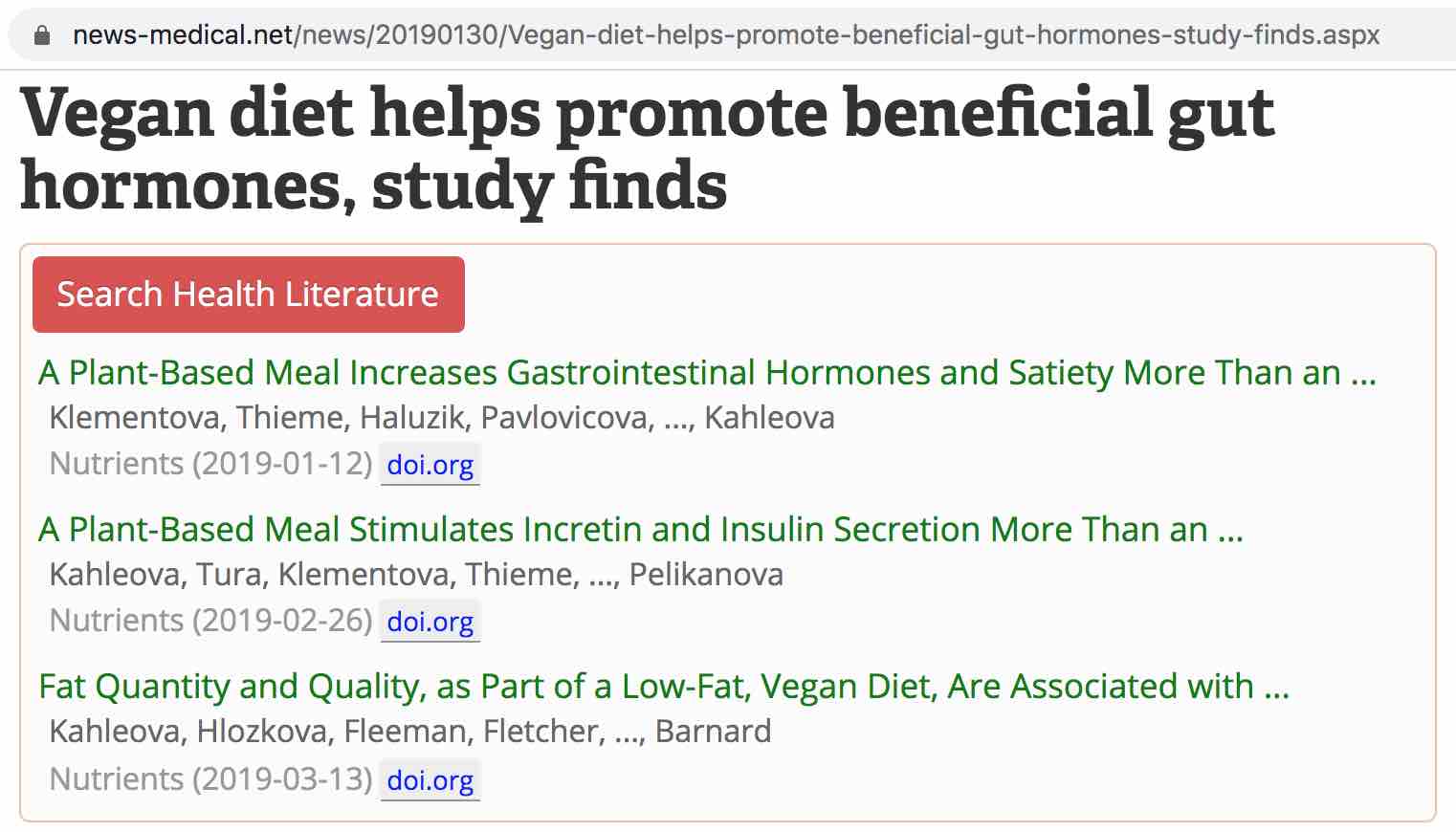}
\caption{Screenshot of a user clicking on an injected button ``Search Health Literature'' on a health news page.
}
\label{figScreenshotOfNews2PubMed}
\end{figure}

\noi {\bf A browser extenstion demonstration.}
We also developed a browser extension to make our linking approach available for public use \cite{wang2021sigirdemo}.
When a reader visits an article on a health news website, 
she will see an injected button {\em Search Health Literature},
underneath the headline of the news article (Fig.~\ref{figScreenshotOfNews2PubMed}).
If the user clicks on the search button, 
the top 3 papers in the search result will be presented inside a box,
where the user can click the title of the paper to view its abstract
or click ``{\tt doi.org}'' to access the paper on its official website.
If none of the matching scores meets a minimum threshold, 
the system will not return any result. 

\section{Conclusion and Future Work}

Linking news to scientific literature is a relatively unexplored area.
In this paper we presented a comprehensive approach to the problem,
from training and test dataset construction, metadata extraction,
to literature search engine development.
Here we highlight the major findings from this study:
(1) we are able to develop deployable machine learning models for journal name identification;
(2) an elastic search engine indexed on 20+ million PubMed records can harness 
the power of various news metadata (journal, author name and affiliation, news title, and even news content);
and (3) our approach can be practically used to link press releases 
posted on \ea to their associated research articles in PubMed,
which facilitates fact checking of press releases and longitudinal trend analysis. 

Our study also poses some challenges and opportunities for future work.
(1) Currently, we focused on the news articles that report on a single research work, usually the latest research findings;
however, some news articles report on multiple studies, especially when introducing a new research area rather than an individual study. In the future we will extend and evaluate our approach to such one-to-many matching cases.
(2) In this study we focused on health news and biomedical literature. In the future we will adapt our approach to domains beyond biomedical and health.
This is now feasible since from April 2021 anyone can freely download the 120 million DOI registration data provided by CrossRef 
\cite{crossref2021cache}
and the similiar amount of abstract or summary data provided by SemanticScholar.

\bibliographystyle{unsrtnat}
\bibliography{references}

\end{document}